\documentclass{appolb}
\usepackage{epsfig}

\begin{document}
\title{Electroweak \& QCD  corrections to \\ 
 Drell Yan processes %
}
\author{G. Balossini, G. Montagna
\address{Dipartimento di Fisica
Nucleare e Teorica, Universit\`a di Pavia,
and INFN, Sezione di Pavia, via A. Bassi 6, 27100, Italy}
\and
C.M. Carloni Calame
\address{INFN, Frascati (Italy) and School of Physics and Astronomy, Southampton
   University, Southampton (UK) }
\and
M. Moretti, M. Treccani
\address{Dipartimento di Fisica, Universit\`a di 
Ferrara, and INFN, Sezione di Ferrara, Italy}
\and
O. Nicrosini, F. Piccinini
\address{INFN, Sezione di Pavia, via A. Bassi 6, 27100, Italy}
\and
A. Vicini
\address{Dipartimento di Fisica, Universit\`a di Milano, and INFN, 
Sezione di Milano, Via~Celoria 16, 20133, Italy}
}
\maketitle

\vskip 60pt 
\begin{center}
Invited talk presented at the 2008 Cracow Epiphany Conference \\
in honour of the $60^{th}$ birthday of Prof. S. Jadach
\end{center}

\begin{abstract}
The relevance of single-$W$ and single-$Z$ production 
processes at hadron colliders is well known: in the present paper  the status of theoretical 
calculations of Drell-Yan processes is summarized  and some results on the combination of electroweak and 
QCD corrections to a sample of observables of the process 
$p p \to W^\pm \to \mu^\pm + X$ at the LHC are discussed. 
The  phenomenological analysis shows that a high-precision knowledge 
of QCD and a careful combination of electroweak and strong 
contributions is mandatory in view of the anticipated LHC experimental accuracy. \\
One of the authors (O.N.) dedicates these notes to Prof. S. Jadach, in honour of his $60^{th}$ birthday and grateful for all that Prof. Jadach taught him during their fruitful collaboration. 
\end{abstract}
\PACS{12.15.Ji, 12.15.Lk, 12.38.-t}
  
\section{Introduction}
Single-$W$ and single-$Z$ production 
processes  are today considered of utmost relevance for the physics studies at contemporary hadron colliders such as Tevatron and the LHC. Actually, 
charged and neutral current Drell-Yan (D-Y) processes, 
{\emph{ i.e.}} $p p\hspace{-8pt}{}^{{}^{(-)}} \to  W\to l \nu_l + X$, 
and $p p\hspace{-8pt}{}^{{}^{(-)}} \to Z/\gamma \to l^+ l^- + X$ 
play a very important role, since they 
have huge cross sections 
(\emph{e.g.} $\sigma(p p \to  W\to l \nu_l + X)$ $\sim$ $20$~nb at LHC and 
about a factor of ten less for 
$\sigma(p p \to Z/\gamma \to l^+ l^- + X)$) and 
are easily detected, given the presence of at least a high $p_\perp$ lepton, 
which to trigger on. 
For this reasons and also because the physics around $W$ and $Z$ 
mass scale is presently known with high precision after the LEP and Tevatron 
experience, D-Y processes will provide standard candles for detector 
calibration during the first stage of LHC running. 
Moreover, single-$W$ as signal by itself 
will allow to perform a precise measurement of the $W$ mass with a 
foreseen final uncertainty of the order of 15~MeV at LHC 
(20~MeV at Tevatron), a very important 
ingredient for precision tests of the Standard Model, when associated with 
a top mass uncertainty of the order of 1-2~GeV. Also, from the 
forward-backward asymmetry of the charged lepton pair in 
$p p \to Z/\gamma \to e^+ e^-$ the mixing angle $\sin^2 \vartheta_W$ could 
be extracted with a precision of $1 \times 10^{-4}$. 
Useful observables for the measurement of the $W$ mass 
are the transverse mass distribution and the charged lepton transverse 
momentum distribution. While the latter is in principle experimentally 
cleaner, the former is less sensitive to the effects of higher order 
radiative corrections affecting the theoretical predictions. 

The few per cent level 
precision in principle achievable in the cross sections motivated a proposal 
to use these observables as luminosity monitor for the LHC. 
Last, single-$W$ and single-$Z$ processes will provide important 
observables for new physics searches: in fact 
the high tail of the $l^+ l^-$ invariant mass and of the $W$ transverse 
mass is sensitive to the presence of extra gauge bosons predicted 
in many extension of the Standard Model, which 
could lie in the TeV energy scale detectable at LHC. 

For the above reasons, it is of utmost importance to predict the 
$W$ and $Z$ observables with as high as possible theoretical 
precision. The sources of uncertainty in the theoretical predictions 
are essentially of perturbative and non-perturbative origin. 
The latter ones comprise the uncertainties related to the parton 
distribution functions and power corrections 
to resummed differential cross sections, 
which will not be discussed here. 
In the following we review the current state-of-the-art 
on the calculation of higher order QCD and electroweak (EW)
radiative corrections and their implementation in simulation tools, and we present some original results about the
combination of QCD and EW corrections to $W$ production
at the LHC. 

\section{Status of theoretical calculations}
\subsection{Higher-order QCD/EW calculations and tools}

In the present section, a sketchy summary of the main computational tools for EW gauge boson 
production at hadron colliders is presented. 
Concerning QCD calculations and tools, the present situation reveals quite a rich structure, 
that includes 
next-to-leading-order (NLO) and next-to-next-to-leading-order (NNLO) 
corrections to $W/Z$ total production rate~\cite{AEM,HvNM}, 
NLO calculations for $W, Z + 1, 2 \, \, {\rm jets}$ 
signatures (available in the codes DYRAD and MCFM)~\cite{GGK,MCFM} , resummation of 
leading and next-to-leading logarithms due to soft gluon 
radiation (implemented 
in the Monte Carlo ResBos)~\cite{BY,resbos},
NLO corrections merged with QCD Parton Shower (PS) 
evolution (in the event generators
MC@NLO and POWHEG)~\cite{MC@NLO, POWHEG}, 
 NNLO corrections to $W/Z$ production in fully differential 
form 
 (available in the Monte Carlo program FEWZ)~\cite{mp,mp1},  
 as well as leading-order multi-parton matrix elements 
generators matched with vetoed PS, such as, for instance, 
ALPGEN~\cite{Alpgen}, MADEVENT~\cite{MadEvent}, HELAC~\cite{Helac} and 
SHERPA~\cite{Sherpa}.

As far as complete ${\cal O}(\alpha)$ EW corrections 
to D-Y processes
are concerned, they have  been computed independently by various 
authors in \cite{dk, bw,ZYK,SANC,CMNV} for $W$ production and 
in \cite{zgrad2, Zykunov2007,HORACEZ, SANCZ} for $Z$ production.
EW tools implementing exact NLO corrections to $W$ 
production are DK \cite{dk}, WGRAD2 \cite{bw}, SANC \cite{SANC} 
and HORACE 
\cite{CMNV}, while ZGRAD2~\cite{zgrad2}, HORACE~\cite{HORACEZ} and SANC~\cite{SANCZ} include the full set of 
${\cal O}(\alpha)$ EW  
corrections to $Z$ production. The predictions of a subset of such 
calculations have been 
compared, at the level of same input parameters and cuts, 
in the proceedings 
of the Les Houches 2005~\cite{LH} and TEV4LHC~\cite{tev4lhc} workshops for $W$ production, 
finding a very satisfactory agreement between the various, 
independent calculations. 
A first set of tuned comparisons for the $Z$ production process has been performed and is available in~\cite{LH2007}. 

From the calculations above, it turns out that NLO EW  
corrections are dominated, 
in the resonant region, by final-state QED radiation containing 
large collinear logarithms
of the form  $\log(\hat{s}/m_l^2)$, where $\hat{s}$ is the squared 
partonic centre-of-mass (c.m.) energy
and $m_l$ is the lepton mass. Since these corrections amount to 
several per cents around the
jacobian peak of the $W$ transverse mass and lepton transverse momentum 
distributions and cause a
significant shift (of the order of 100-200~MeV) in the extraction of the 
$W$ mass $M_W$ at the Tevatron, 
the contribution of higher-order corrections due to multiple photon 
radiation from the 
final-state leptons must be taken into account in the theoretical 
predictions, in view of
the expected precision (at the level of 15-20 MeV) in the $M_W$ 
measurement at the LHC.  The contribution
 due to multiple photon radiation has been computed, by means of a QED PS
 approach, in \cite{CMNTW} for $W$ production and
 in \cite{CMNTZ} for $Z$ production, and implemented in the event generator HORACE. 
 Higher-order QED contributions to $W$ production have been calculated independently
 in \cite{winhac} using the YFS exponentiation, and are available in the generator
 WINHAC. They have been also computed in the collinear approximation, within the structure functions approach, in~\cite{DK2007}. 

It is worth noting that, for what concerns the precision
 measurement of $M_W$, the shift induced by higher-order QED corrections is about 
 10\% of that caused by one-photon emission and of opposite sign, as shown in 
 \cite{CMNTW}. Therefore, such an effect is non-negligible in view of the aimed accuracy 
 in the $M_W$ measurement at the LHC. 

A further important phenomenological feature of EW corrections is that, in the 
region important for new physics searches (i.e. where the $W$ transverse mass is much
larger than the $W$ mass or the invariant mass of the final state leptons is much larger
than the $Z$ mass), the NLO EW effects become large (of the order of 
20-30\%) and negative, due to the appearance of EW Sudakov logarithms 
$\propto - (\alpha/\pi) \log^2 ({\hat s}/M_V^2)$, $V = W,Z$ \cite{dk,bw,CMNV,zgrad2,Zykunov2007,HORACEZ}. 
Furthermore, in this region, weak boson emission processes 
(e.g. $pp\to e^+\nu_eV + X$), 
that contribute at the same order in perturbation theory, can partially cancel the large Sudakov 
corrections, when the weak boson $V$ decays into unobserved $\nu\bar{\nu}$ or
jet pairs, as recently shown in \cite{baurw}.

\subsection{Combination of EW and QCD corrections}

In spite of this detailed knowledge of higher-order EW and QCD 
corrections, the combination of their effects is presently under investigation. 
Some attempts have been explored in the literature~\cite{cy, ward2007, jadach2007}. 
Here our approach will be discussed in some detail. 

A first strategy  for the combination of EW and QCD 
corrections consists in the 
following formula

\begin{eqnarray}
\left[\frac{d\sigma}{d\cal O}\right]_{{\rm QCD} \& {\rm EW}} = 
\left\{\frac{d\sigma}{d\cal O}\right\}_{{\rm MC@NLO}}
+\left\{\left[\frac{d\sigma}{d\cal O}\right]_{{\rm EW}} - 
\left[\frac{d\sigma}{d\cal O}\right]_{{\rm Born}} \right\}_{{\rm HERWIG\, \,  PS}}
\label{eq:qcd-ew}
\end{eqnarray}
where ${d\sigma/d\cal O}_{{\rm MC@NLO}}$ stands for the prediction of the 
observable ${d\sigma/d\cal O}$ 
as  obtained  by means of MC@NLO, 
%
%
${d\sigma/d\cal O}_{{\rm EW}}$ is the HORACE
prediction for the EW corrections to the ${d\sigma/d\cal O}$ observable, 
and ${d\sigma/d\cal O}_{{\rm Born}}$ is the lowest-order result 
for the observable of interest. The label {{\rm HERWIG PS} in the second term in r.h.s. 
of eq. (\ref{eq:qcd-ew}) means that EW corrections are convoluted with QCD PS 
evolution through the HERWIG event generator, in order to (approximately) include
mixed ${\cal O}(\alpha \alpha_s)$ corrections and to obtain a more realistic 
description of the observables under study. In eq.~(\ref{eq:qcd-ew}) the infrared part of QCD corrections is factorized, whereas the infrared-safe matrix element residue is included in  an additive  form. It is otherwise possible to implement a fully factorized combination (valid for infra-red safe observables) as follows: 

\begin{eqnarray}
\left[\frac{d\sigma}{d\cal O}\right]_{{\rm QCD} \otimes {\rm EW}} = & &
\left( 1 + \frac{
\left[{d\sigma} / {d\cal O}\right]_{\rm MC@NLO} - \left[{d\sigma}/{d\cal O}\right]_{\rm HERWIG\, \, PS}
}{
\left[{d\sigma}/{d\cal O}\right]_{\rm Born}
}
\right) \times \nonumber \\
& & \times 
\left\{ 
\frac{d\sigma}{{d\cal O}_{\rm EW}}
\right\}_{{\rm HERWIG\, \,  PS}} , 
%
\label{eq:qcd-ew-factor}
\end{eqnarray}
where the ingredients are the same as in eq.~(\ref{eq:qcd-ew}) but also the QCD matrix element residue in now factorized.  Eqs.~(\ref{eq:qcd-ew}) and ~(\ref{eq:qcd-ew-factor}) have the very same ${\cal O}(\alpha)$ and ${\cal O}(\alpha_s)$ content, differing by terms at the order $\alpha \alpha_s$. Their relative difference has been checked to be of the order of a few per cent in the peak region, and can be taken as an estimate of the uncertainty of QCD \& EW combination. 
\begin{center}
\begin{figure}[h]
\hskip 16pt \includegraphics[width=5.5cm]{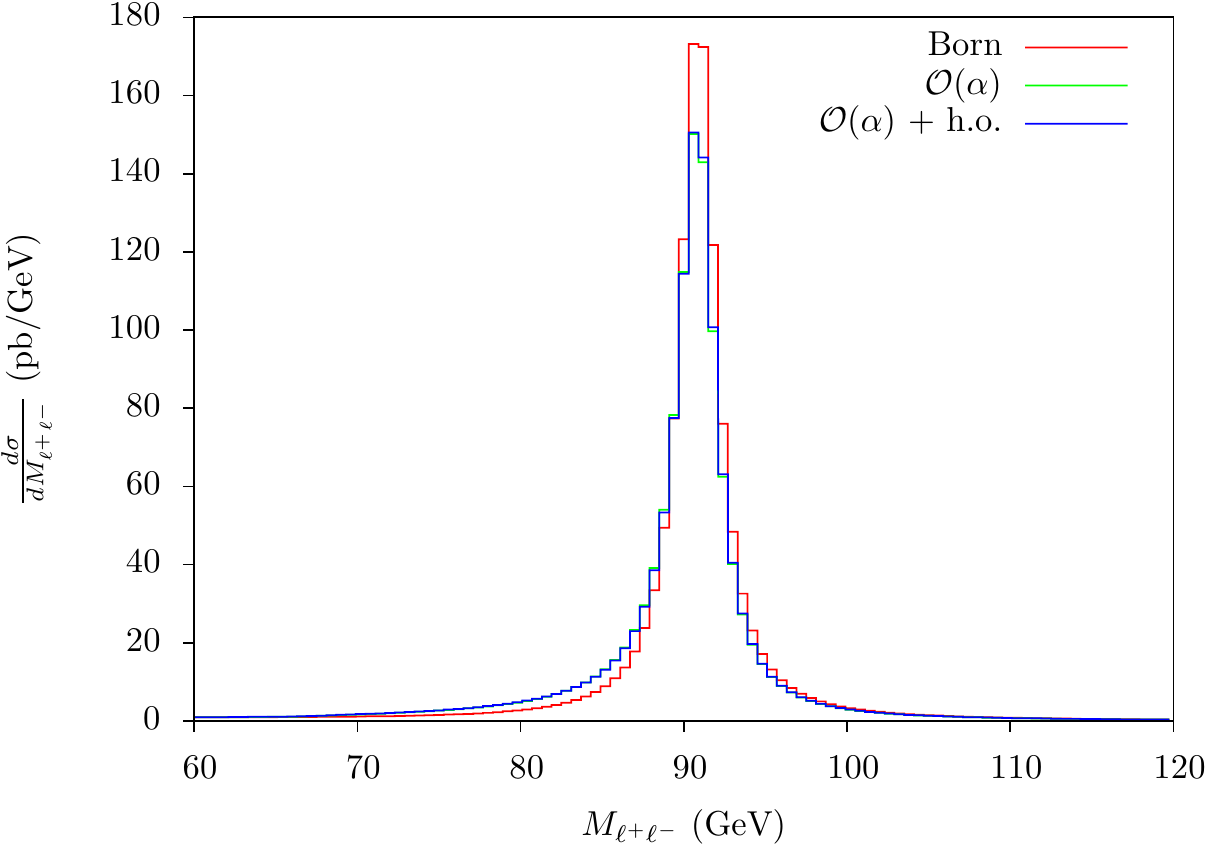}~\includegraphics[width=5.5cm]{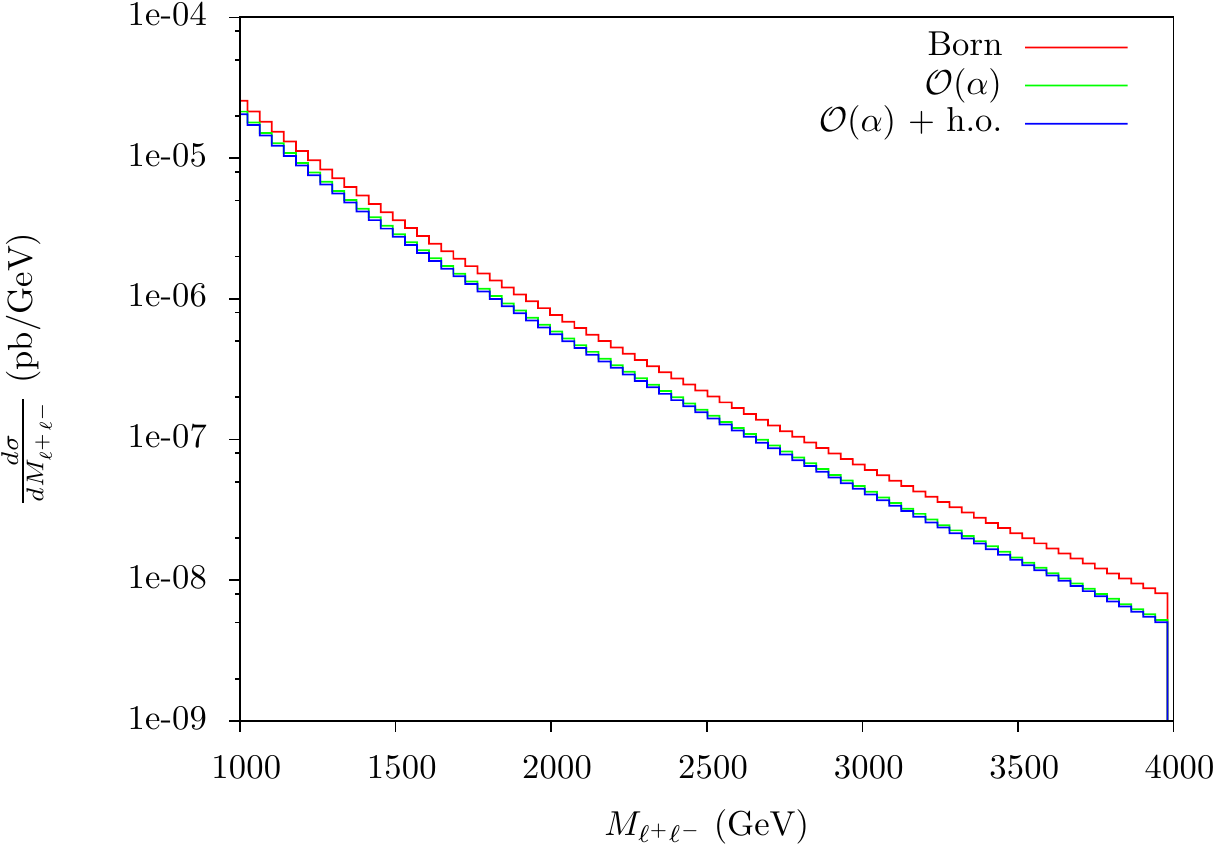}

\hskip 16pt \includegraphics[width=5.5cm]{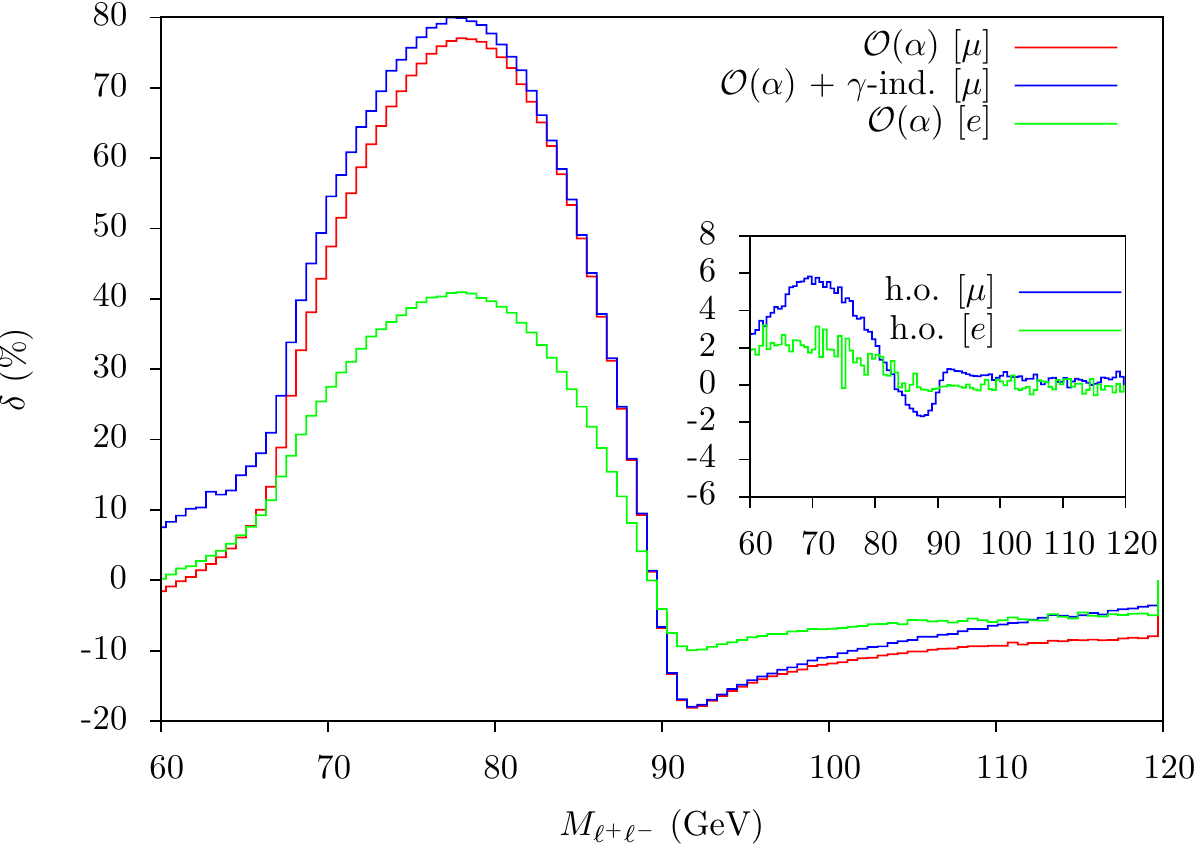}~\includegraphics[width=5.5cm]{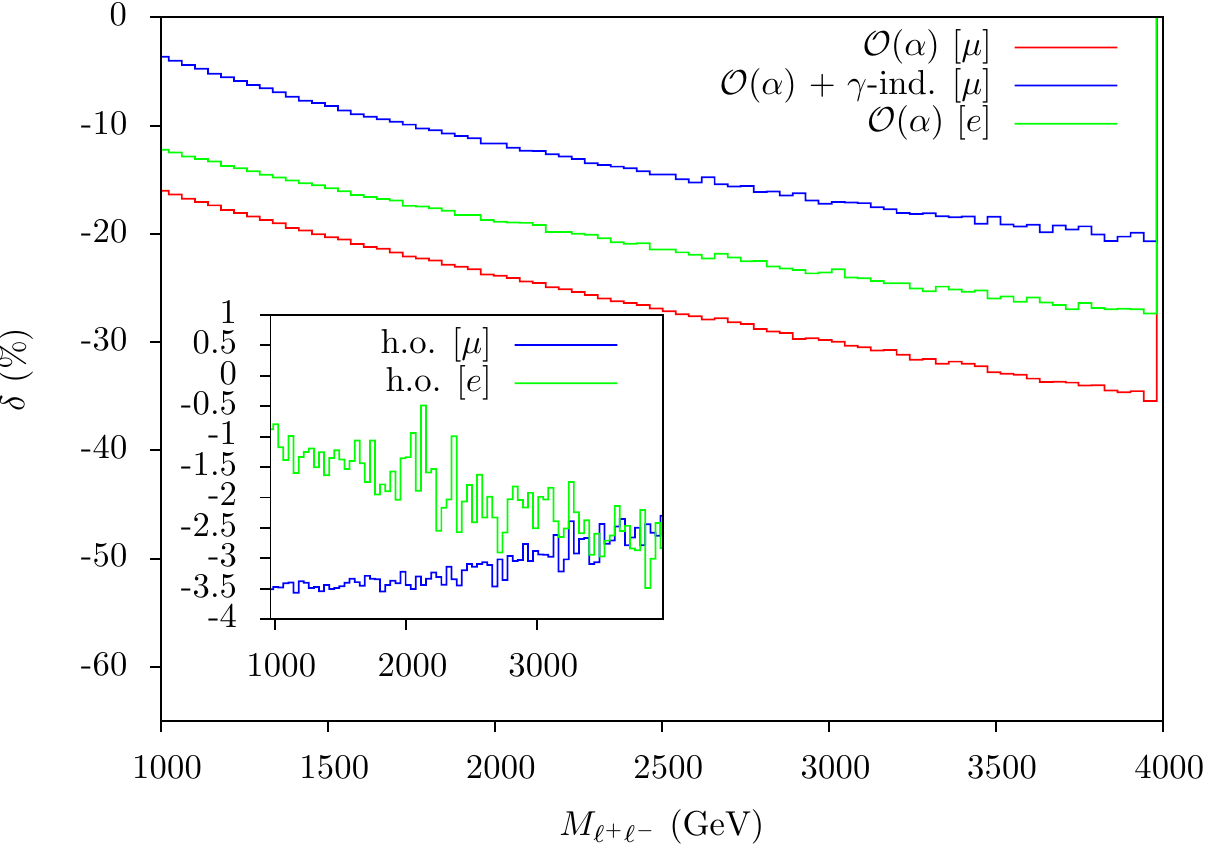}
\caption{Upper panel: HORACE predictions for the $Z$ invariant mass distribution around the peak (left) and in the high tail (right). Lower panel: relative effect of EW corrections.}
\label{Zfig}
\end{figure}
\end{center}

\section{Numerical results}

In order to assess the phenomenological relevance of radiative corrections to D-Y processes, we show the effect of purely EW corrections to $Z$-boson production at the LHC ($\sqrt{s}$ = 14~TeV) in Fig.~\ref{Zfig}. Input parameters, cuts and lepton identification criteria can be found in ref.~\cite{HORACEZ}. The set of PDFs used in our study is MRST2004QED \cite{mrst04qed}. As can be seen, EW corrections give huge contributions around the $Z$ peak, dominated by photonic final state radiation. There are important corrections in the hard invariant mass tail, mainly due to combined photonic and Sudakov effects. Multiple photon corrections are at the some per cent level. 

As far as the combination of QCD \& EW corrections is concerned, we study, for definiteness, the production process $p p \to W^\pm \to 
\mu^\pm + X$ at the LHC, 
imposing the cuts shown in Tab.~\ref{tab:lhc}, where  $p_{\perp}^{\mu}$ and
$\eta_\mu$ are the transverse momentum and the pseudorapidity of the muon, $\rlap{\slash}{\! E_T} $
is the missing transverse energy, which we identify with the transverse momentum of 
the neutrino, as typically done in several phenomenological studies. For 
set up b., a severe cut on the $W$ transverse mass $M_\perp^W$ is 
superimposed to the cuts of set up a., in order to isolate the region 
of the high tail of $M_T^W$, which is interesting for new physics searches. 
The QCD 
factorization/renormalization scale and the analogous QED scale (present 
in MRST2004QED) are chosen to be equal, as usually done in the literature \cite{dk,bw,CMNV}, and fixed at $\mu_R = \mu_F = \sqrt{p_{\perp W}^2 + M_{l\nu_l}^2}$, where $M_{l\nu_l}$ is the $W$-boson invariant mass. 

\begin{table}[h]
\begin{center}
\begin{tabular}{|c|}
\hline
{\large LHC}\\
\hline
$\, \,$ a. $\,$ $p_{\perp}^{\mu} \geq$~25 GeV $\, \, $ $\rlap{\slash}{\! E_T}  \geq$~25 GeV and 
$|\eta_\mu|< 2.5$\\
\hline
$\!  \! \! \! \! \! \! \! \! \! \! \! \! \! \! \! \!$ b. the cuts as above $\, \, $  $\oplus$ $\, \,$ $M_\perp^W \geq 1$~TeV\\
\hline
\end{tabular}
\caption{Selection criteria imposed for the numerical simulation of single-$W$ 
production process at the LHC.}
\label{tab:lhc}
\end{center}
\end{table}

\begin{center}
\begin{figure}[h]
\includegraphics[width=5.9cm]{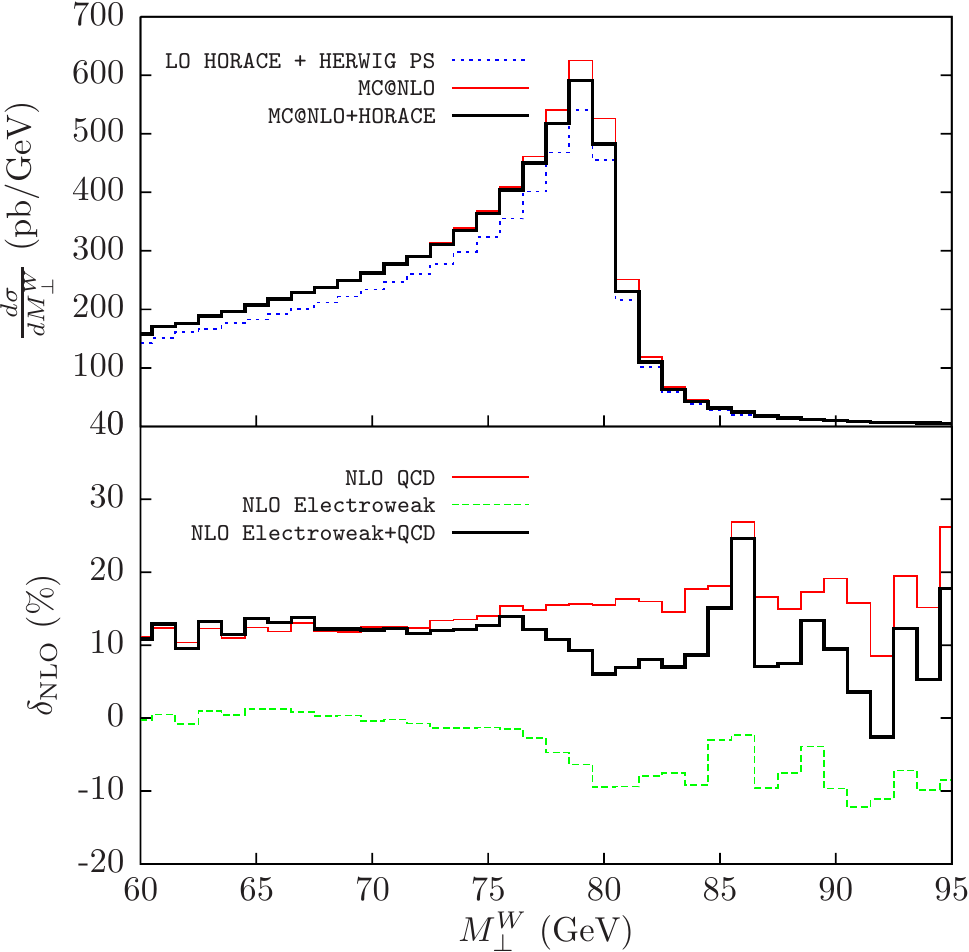}~\hskip 8pt\includegraphics[width=5.9cm]{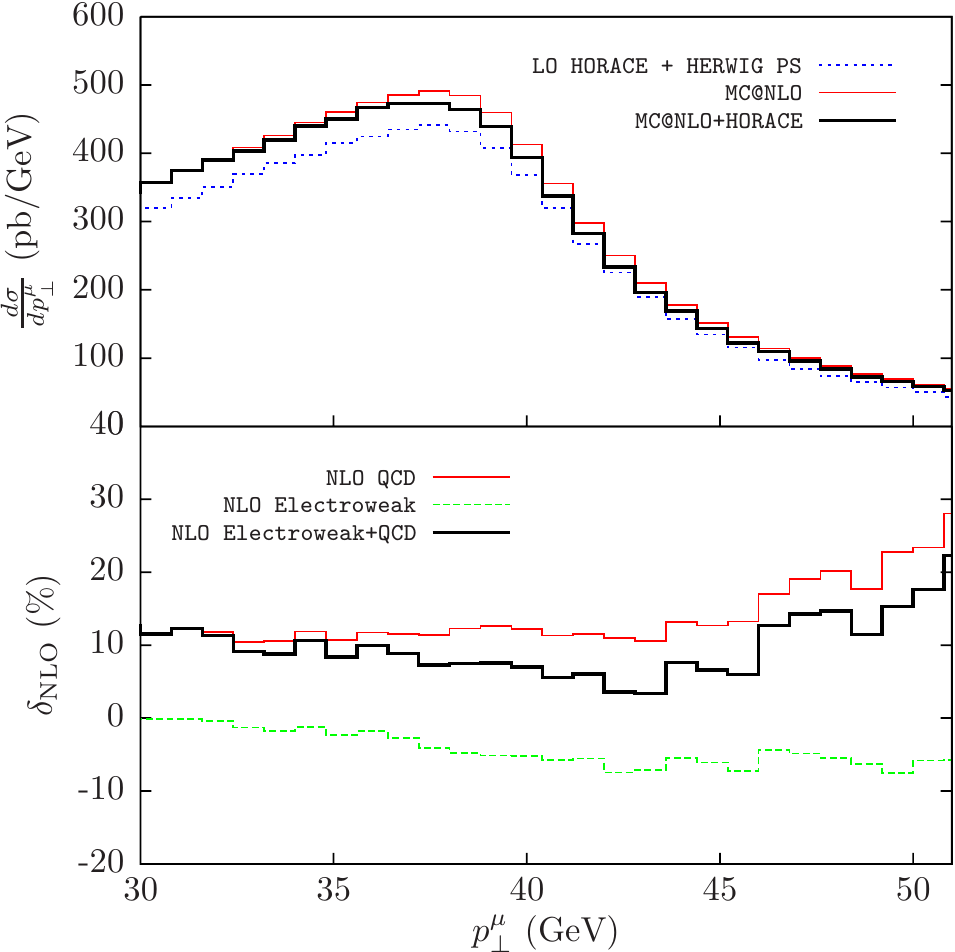}
\caption{Upper panel: predictions of MC@NLO, MC@NLO+HORACE and leading-order  HORACE+HERWIG 
PS for the $M_\perp^W$ (left) and $p_\perp^\mu$ (right) distributions at the LHC, according to the cuts of set up a. Lower panel: relative effect of QCD and EW corrections, and their sum, for the 
corresponding observables in the upper panel.}
\label{peak}
\end{figure}
\end{center}

In order to avoid systematic theoretical effects, all the generators under consideration have been properly tuned to reproduce the same LO/NLO results. 
A sample of our numerical results is shown in Fig. \ref{peak} for the $W$ transverse mass  $M_\perp^W$ and muon transverse momentum  $p_{\perp}^{\mu}$ distributions according to set up a. 
of Tab.~\ref{tab:lhc}, and in Fig. \ref{offpeak} for the
same distributions according to set up b. In each figure, the upper panels show the 
predictions of the generators MC@NLO and MC@NLO + HORACE interfaced to 
HERWIG PS (according to eq.(\ref{eq:qcd-ew})), in 
comparison with the leading-order result by HORACE convoluted with HERWIG
shower evolution. The lower panels illustrate the relative effects of  the matrix element residue of NLO QCD and full 
EW corrections, as well as their sum, that can be obtained by appropriate 
combinations of the results shown  in the upper panels. From Fig. \ref{peak} it can be seen that  QCD corrections are positive around the 
jacobian peak and tend to compensate the effect due to EW corrections. Therefore, 
their interplay is crucial for a precise $M_W$ extraction at the LHC and their combined 
contribution can not be accounted for in terms of a pure QCD PS approach, 
as it can be inferred from the comparison of the predictions of MC@NLO versus 
the leading-order result by HORACE convoluted with HERWIG PS.

The interplay between QCD and EW corrections in the region interesting for 
new physics searches, i.e. in the high tail of $M_\perp^W$ and $p_\perp^\mu$ distributions,
is shown in Fig. \ref{offpeak}. For both $M_\perp^W$ and $p_\perp^\mu$ 
NLO QCD corrections are positive and largely cancel the 
negative EW Sudakov logarithms. Therefore, a precise
normalization of the SM background to new physics searches 
necessarily requires the simultaneous control of QCD and 
EW corrections. 

\begin{center}
\begin{figure}[h]
\includegraphics[width=5.9cm]{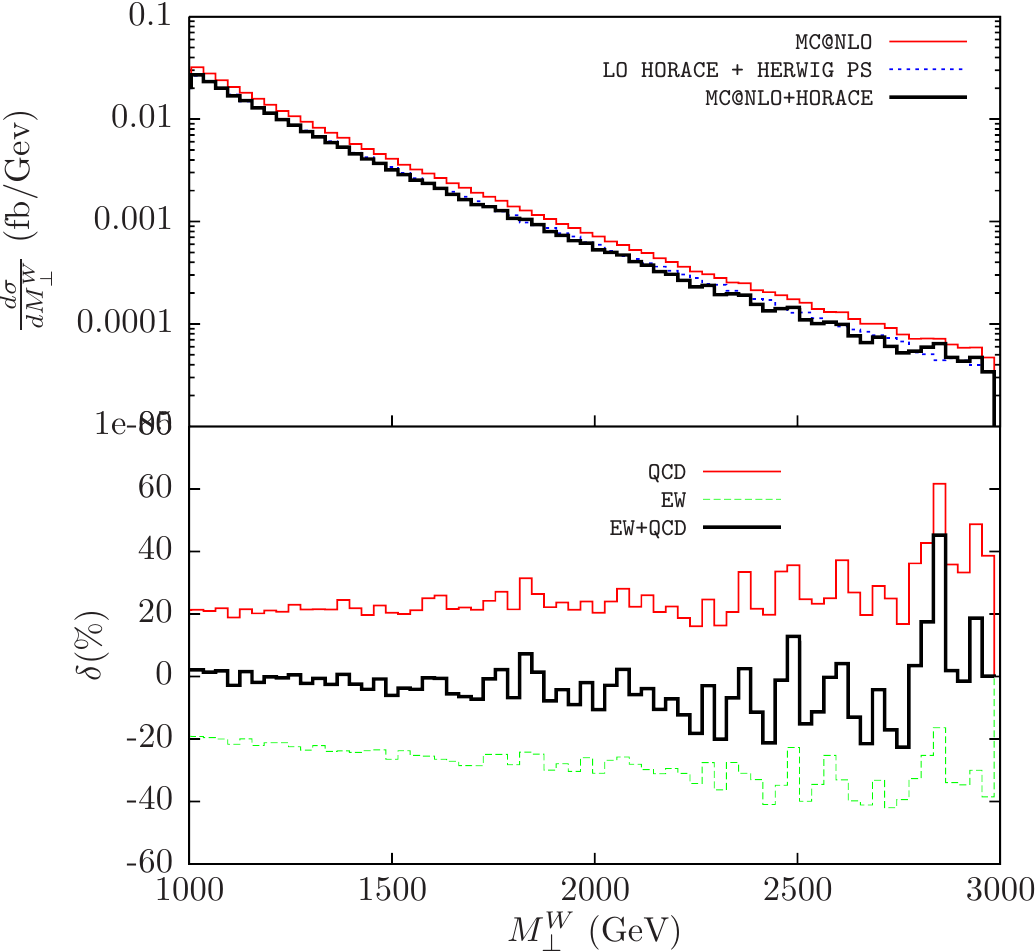}~\hskip 8pt\includegraphics[width=5.9cm]{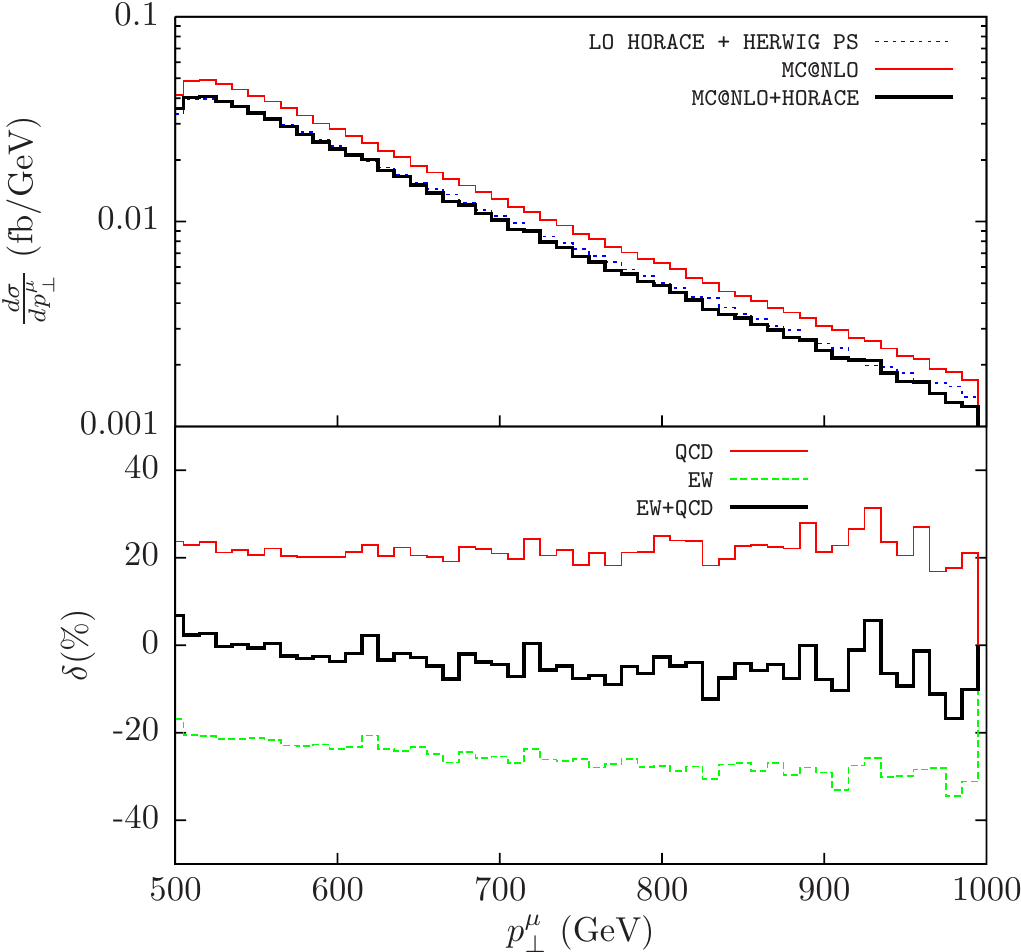}
\caption{The same as Fig.~\ref{peak} according to the cuts of set up b.}
\label{offpeak}
\end{figure}
\end{center}
\section{Conclusions}
During the last few years, there has been a big effort towards high-precision
predictions for D-Y-like processes, addressing the calculation of higher-order QCD and EW corrections. Correspondingly, precision computational tools have been developed to keep under control theoretical systematics in view of the future measurements at the LHC.

 We presented some results about EW and 
QCD corrections to a sample of observables of the $Z$ and $W$ production processes at the LHC. 
Our investigation shows that a high-precision knowledge of QCD and a careful combination of electroweak and strong contributions is mandatory in view of the anticipated experimental accuracy. We plan, however,
to perform a more complete and detailed phenomenological study, including the predictions of other 
QCD generators and considering further observables of interest for the many facets of the
$W/Z$ physics program at the LHC.  

\vskip 24pt\noindent
\leftline{\bf Acknowledgements}
O. Nicrosini would like to thank the organizers 
for their kind invitation and warm hospitality. He is also grateful to J.H. Kh\"un for discussions about factorization.

\end{document}